\documentclass{aa} 
\usepackage[american]{babel}
\usepackage{graphicx} 
\usepackage{txfonts} 
\usepackage{natbib} 
\bibpunct{(}{)}{;}{a}{}{,} 

\begin{document}
\def\lsim{\,\lower2truept\hbox{${< \atop\hbox{\raise4truept\hbox{$\sim$}}}$}\,}
\def\gsim{\,\lower2truept\hbox{${> \atop\hbox{\raise4truept\hbox{$\sim$}}}$}\,}
\title{A numerical code for the solution of the Kompaneets equation 
in cosmological context}

     \author{P.~Procopio\inst{1,2}  
         \and  C.~Burigana\inst{1,2}}
   \institute{INAF/IASF-BO, Istituto di Astrofisica Spaziale e Fisica 
Cosmica di Bologna,
    Via Gobetti 101, I-40129, Bologna, Italy 
    \and Dipartimento di Fisica, Universit\`a degli Studi di Ferrara, 
Via Saragat 1, I-44100 Ferrara, Italy}
%

\abstract
{The cosmic microwave background (CMB) spectrum probes physical
processes and astrophysical phenomena occurring at various epochs
of the Universe evolution.
Current and future CMB absolute temperature experiments
are aimed to the discovery of the very small distortions
such those associated to the cosmological reionization process
or that could be generated by different kinds of earlier processes.
The interpretation of future data calls for a continuous improvement in
the theoretical modeling of CMB spectrum.}
{In this work we describe the fundamental approach
and, in particular, the update to recent NAG versions
of a numerical code, KYPRIX,
specifically written for the solution of the Kompaneets
equation in cosmological context, first implemented in the years
1989-1991, aimed at the very accurate computation
of the CMB spectral distortions
under quite general assumptions.}
{We describe the structure and the main subdivisions of
the code and discuss the most relevant aspects of its technical
implementation.}
{We present some of fundamental tests we carried out to
verify the accuracy, reliability, and performance
of the code.}
{All the tests done demonstrates the reliability and versatility
of the new code version and its very good accuracy and
applicability to the scientific analysis of current CMB
spectrum data and of much more precise measurements that will be
available in the future.
The recipes and tests
described in this work can be also useful to implement accurate
numerical codes for other scientific purposes using
the same or similar numerical libraries or to verify the validity
of different codes aimed at the same or similar problems.}

\keywords
{Cosmic microwave background -- Radiation mechanisms 
-- Radiative transfer -- Scattering}
%
\titlerunning{A code for the Kompaneets equation in cosmology}
\maketitle


\section{Introduction}
\label{sec:intro}

   \let\thefootnote\relax\footnotetext{The address to which the proofs have to be sent is: \\
   Carlo Burigana\\
   INAF-IASF Bologna,
   Via Gobetti 101, I-40129, Bologna, Italy\\
   fax: +39-051-6398724\\
   e-mail: burigana@iasfbo.inaf.it}

The CMB spectrum emerges from the thermalization redshift,
$z_{therm} \sim 10^6 - 10^7$,
with a shape very close to a Planckian one,
owing to the tight coupling between radiation and matter through
Compton scattering and photon production/absorption processes,
radiative Compton and bremsstrahlung. These processes
were extremely efficient at early times
and able to re-establish a blackbody (BB) spectrum
from a perturbed one
on timescales much shorter than the expansion time (see e.g. \cite{DD77}).
The value of $z_{therm}$ 
\citep{BDD91a}
depends on the baryon density parameter,
$\Omega_b$, and the Hubble constant, $H_0$, through the product
$\widehat \Omega=\Omega_b (H_{0}/50)^2$ ($H_0$ expressed in Km/s/Mpc).

On the other hand, physical processes occurring at redshifts $z < 
z_{therm}$
may leave imprints on the CMB spectrum.
Therefore, the CMB spectrum carries crucial informations on physical
processes occurring during early cosmic epochs
(see e.g. \cite{DB93} and references therein)
and the comparison between models of CMB spectral distortions
and CMB absolute temperature measures can constrain the
physical parameters of the considered dissipation processes
\citep{BDD91b}.

The timescale for the achievement of
kinetic equilibrium between radiation and matter
(i.e. the relaxation time for the photon spectrum), $t_C$, is
\begin{equation}
t_C=t_{\gamma e} {m_e c^{2}\over {kT_e}} \simeq 
4.5 \times 10^{28}
( T_{0}/2.7\, K ) ^{-1} \phi ^{-1} \widehat \Omega^{-1}
(1+z )^{-4} \sec \, ,
\label{eq_t_c}
\end{equation}
%
where $t_{\gamma e}= 1/(n_e \sigma _T c)$ is the photon--electron 
collision
time, $\phi = (T_e/T_r)$, $T_e$ and $T_r=T_{0}(1+z)$ being
respectively the electron and the CMB radiation temperature;
$kT_e/m_e c^2$ (being $m_e$ the electron mass) is the mean fractional 
change of photon energy in a scattering
of cool photons off hot electrons, i.e. $T_e \gg T_r$;
$T_0$ is the present radiation temperature related
to the present radiation energy density by $\epsilon _{r0}={\rm a}T_0^4$
(here ${\rm a}=8 \pi I_3 k^4/(hc)^3$, $I_3=\pi^4/15$).
A primordial helium abundance of 25\% in mass is assumed in these
numerical estimates.
It is useful to introduce the dimensionless time variable $y_e(z)$
defined by
\begin{equation}
y_e(z) = \int^{t_0}_{t} {dt \over t_C}
=\int^{1+z}_{1} {d(1+z) \over 1+z} {t_{exp}\over t_C} \, ,
\label{eq_y_e}
\end{equation}
where $t$ is the time, $t_0$ is the present time, and
$t_{exp}=1/{H}=1/[(da/dt)/a]$ is the expansion time, $a=1/(1+z)$ being 
the cosmic scale factor normalized to the present time.

As particular cases,
by neglecting the cosmological constant (or dark energy)
contribution we have
%
\begin{equation}
t_{exp} \simeq   6.3\times 10^{19} \left({T_0 \over 2.7\, K}\right)^{-2}
(1+z)^{-3/2} \times
\label{eq_t_exp}
\end{equation}
$$
 \left[\kappa (1+z) + (1+z_{eq})
-\left({\Omega_{nr} -1 \over \Omega_{nr}}\right)
\left({1+z_{eq} \over 1+z}\right) \right]^{-1/2}
\sec \, ,
$$
where $z_{eq} = 1.0\times 10^4 (T_{0}/2.7\, K)^{-4}\widehat{\Omega}_{nr}$
is the redshift of
equal non relativistic matter and photon energy densities
and $\kappa = 1 + N_\nu (7/8)
(4/11)^{4/3}$, $N_\nu$ being the number of relativistic, 2--component,
neutrino species (for 3 species of massless neutrinos, $\kappa \simeq 
1.68$),
takes into account the
contribution of relativistic neutrinos to the dynamics of the
Universe$^1$\footnote{$^1$Strictly speaking the present ratio of 
neutrino to photon energy densities, and hence the value of $\kappa$, is 
itself a function of the amount of energy dissipated. The effect, 
however, is never very important and is negligible
for very small distortions.}, while
assuming $\Omega_K=0$, $\Omega_{\Lambda}=1-\Omega_{nr}$,
and neglecting the radiation energy density,
as possible at relatively low redshifts,
we have
\begin{equation}
t_{exp} \simeq (1/H_0)
\left[\Omega_{nr} (1+z)^3 + 1-\Omega_{nr} \right]^{-1/2}
\sec \, ,
\label{eq_t_exp_lowz}
\end{equation}
where $1/H_0 \simeq 3.1 \times 10^{17} h^{-1}$~sec ($h=H_0/100$).

The time evolution
of the photon occupation number, $\eta (\nu,t)$,
under the combined effect of Compton scattering 
and of photon production processes, namely radiative Compton (RC) 
\citep{Gould},
bremsstrahlung (B) 
\citep{KL61, RL79}, 
plus other possible photon emission/absorption contributions
(EM)$^2$\footnote{$^2$A process that in principle should be included  
is the cyclotron emission. 
On the other hand, 
for realistic values of cosmic magnetic field, the 
cyclotron process never plays
an important role for (global) CMB spectral distortions when ordinary 
and stimulated emission
and absorption are properly taken into account and CMB realistic 
distorted spectra are considered \citep{zizzo}. 
In fact, the cyclotron term may be 
significant, in the case of
deviations of $\eta$ from the BB distribution at the electron 
temperature, only at very long wavelengths,
corresponding to the cyclotron frequency, where, during the formation of 
a spectral distortion, FF
and RC are able to keep $\eta$ extremely close to the BB equilibrium.},
is well described by 
the complete Kompaneets equation 
\citep{Kompa,BDD95}:

\begin{equation}
{\partial \eta \over \partial t}
={1 \over \phi} {1 \over t_C} {1 \over x^2}
{\partial \over \partial x}
\left[{ x^4
\left[{\phi {\partial \eta \over \partial x}
+\eta (1+\eta)}\right] }\right]
\label{kompa}
\end{equation}
$$
+\left[{\partial \eta \over \partial t}\right]_{RC}
+\left[{\partial \eta \over \partial t}\right]_{B}
+\left[{\partial \eta \over \partial t}\right]_{EM} \, .
$$

\noindent
This equation is coupled to the time differential
equation governing the electron temperature evolution for an arbitrary
radiation spectrum in the presence of Compton scattering, energy losses
due to radiative Compton and bremsstrahlung, adiabatic
cooling, plus possible external heating sources, $q=a^{-3}(dQ/dt)$,

\begin{equation}
{dT_e \over dt}={T_{eq,C}-T_e \over (27/28)t_{e\gamma}}
-{2T_e \over t_{exp}} +\left[{dT_e \over dt}\right]_{RC,B}
+{(32/27)q \over 3n_ek} \, ;
\label{eq_t_e}
\end{equation}
here
$T_{eq,C}=[h\int\eta (1+\eta) \nu^4d\nu] /
[4k\int\eta\nu^3d\nu] \,$ is the Compton equilibrium
electron temperature 
\citep{peyraud, ZL70}, 
$t_{e\gamma}=3m_ec/4\sigma_T\epsilon_r$, 
$\epsilon_r \simeq \epsilon_{r0}(1+z)^4$
being $\epsilon_{r0} = aT_{0}^4$
the radiation energy density today, and
$x = h\nu_{0}/kT_{0}=h\nu_{0}(1+z)/kT_{0}(1+z)$
is a dimensionless, redshift independent, frequency
($\nu_{0}$ being the present frequency).

\section{Setting up the problem}
\label{sec:set-up}

Partial differential linear equations are divided in three classes:
elliptic, parabolic and hyperbolic.  The Kompaneets equation is a
parabolic partial differential equation \citep{tricomi}.  Solutions to
this equation under general conditions have to be searched
numerically, because it is impossible to find analytical solutions
that accurately take into account the many kinds of cosmological
scenarios and the great number of relevant physical processes.  The
numerical code KYPRIX was written to overcome
the limited applicability of analytical solutions and to get a precise
computation of the evolution of the photon distribution function for a
wide range of cosmic epochs and for many cases of cosmological
interest 
\citep{BDD91a}.
KYPRIX makes use of the NAG libraries \citep{NAG_lib}.

Besides these libraries, a lot of numerical algorithms 
are used in the code: we used some of the routines available 
to the scientific community, but often we did write 
routines dedicated to a specific task.
Among the formers, the D03PCF 
routine of the current version of the NAG release
has been used to 
reduce the Kompaneets equation into a system of ordinary 
differential equations 
\citep{dew_walsh,berzins_etal,berzins,skeel_berzins}.
The D03PGF routine used in the first versions of KYPRIX
is no longer available (see also Sect.~\ref{sec:tec_spec}). 
The two codes work adopting the same numerical framework 
or, in other words, the D03PCF routine
of the current NAG release 
corresponds to the D03PGF routine of the NAG release used in 
the first versions of KYPRIX.  
On the other hand, they come from 
different technical implementations
and exibit remarkable differences in several aspects.
The main differences between the two routines, 
their usage, and the corresponding implications
for the code KYPRIX will be described in this work.
In order to use the D03PCF routine, 
we have to put the Kompaneets equation in the form 
\begin{equation}
\sum_{j=1}^{NPDE} P_{i,j} \frac{\partial U_j}{\partial Y} + Q_i 
= X^{-m} \frac{\partial}{\partial X}\Big(X^m R_i \Big)\, ;
\label{pdedef}
\end{equation}
where $Y$ is the time variable and $X$ the spatial variable.
Here $i=1$ and $NPDE=1$.
$P_{i,j},Q_i,R_i$ depend on $x,t,U,\partial U/\partial x$, 
the vector $U$ is defined as immediately below. 
In our case, $P_{i,j}=1$  and $m=0$ (Cartesian coordinates). 
Note that $P_{i,j},Q_i,R_i$ do not depend on 
$\partial U/\partial t$.

The variables that enter in this equation 
are introduced and used in 
logarithmic form, 
to have a good and essentially uniform accuracy of the solution
in the whole considered frequency range.
They are $X$=log($x$) and $U$=log($\eta$).
Hereafter, we will use the variables $X, U, Y$ for 
what concerns the informatic aspect of the problem, 
keeping the use of the variables 
$x, \eta, y$ for considerations directly linked to physical aspects.
The function $R_i$ is determined only 
by the inverse Compton term while the other physical processes, 
i.e. at least Compton scattering, bremsstrahlung, 
and radiative Compton, are included in the function $Q_i$. 
In order to reduce Eq.~(\ref{pdedef}) into a system 
of ordinary differential equations, the D03PCF routine uses the method
of lines: the right member 
of Eq.~(\ref{pdedef}) is discretized, reducing the calculation of partial 
derivatives in terms of finite values of the 
solution vector $U$ at all the points of the $X$ axis grid. Spatial 
discretization is made by the method of finite 
differences \citep{MG}. 
The resulting system of ordinary 
differential equations is solved using a 
differentiation formula method.  
The choice of the time parameter was driven by the need to 
have a very simple form of the Kompaneets equation. Finally, 
a ``temperature independent'' (time) Comptonization parameter 
\begin{equation}
Y = y(t) = \int\frac{dy_e}{\phi} = 
\int_{t_i}^t n_e\sigma_{T}c\frac{kT_r}{mc^2}dt'\,,
\label{eq_y_t}
\end{equation}
has found to be particularly advantageous
\citep{BDD91a}.

\subsection{Boundary conditions}
\label{sec:bndry}

Integrating equations of the type of Eq.~(\ref{pdedef})
means to calculate the time 
evolution of the function $U(X,Y)$, 
for a given initial condition $U(X,0)$ 
(in fact, the problem is also called ``problem at initial conditions''). 
Numerically, the derivatives of $U$ are replaced by finite differences 
between values of $U$ computed at a 
grid of points (in $X$) and the differential equation is replaced by a 
system of more simple equations. 
However, in presence of the only initial condition, this system is singular 
\citep{press}. For 
this reason, resolving partial differential parabolic equations 
needs boundary conditions: the problem is 
at initial values for the $Y$ variable and at the boundary values 
for the $X$ variable. In general,  
boundary conditions mean additional 
relations written to be joined to the system derived from the 
discretization to finite differences.\\
Therefore, a good statement of the problem needs the definition of 
appropriate boundary conditions.
The capability of a refresh of these conditions along the 
integration in time leads more stability to the solution evolution 
because of the evolution of the radiation field. 
Thanks to the opportunity of having the correct value of $\phi$ for 
each time step, the update of the boundary conditions 
can be physically motivated (see also Sect.~\ref{sub:te}).
The limits of the frequency range considered are: 
$X_{min}=\textrm{log}(x_{min}) = -4.3$ e $X_{max}=\textrm{log}(x_{max}) = 1.7$. 
Of course, we want a solution of the 
Kompaneets equation over all the frequency 
range where it is possible to measure the CMB.
Also, we need to consider a frequency range large enough to contain,
in practice, all the energy density of the cosmic radiation field.

The frequency range is so wide for the other two reasons described below. 

During the time evolution, 
some spurious oscillations of the solution at 
points close to the boundaries may appear 
(these effects, that 
could also occur independently of the need of refreshing $\phi$ -- for 
example for cases at constant $\phi$ --,
may be partially amplified if, for computational reasons discussed 
in the following, the necessary refresh of the electronic 
temperature is not made for every time step).
Fixing the frequency integration range limits far from the interval 
where we are interested 
to compute the photon distribution function
allows to prevent 
the ``contamination'' of the solution 
by this possible spurious oscillations in the frequency range
of interest. 

Since we can generally assume 
that a Planckian spectrum at $x_{min}$ is formed 
before recombination in a timescale shorter than the 
expansion time and, on the contrary, at $x_{max}$ 
the shape of the spectrum is unknown, 
it has been implemented in the code the possibility to adopt 
a particular case of Neumann boundary conditions: 
the requirement that the current density, in the frequency space, 
is null at the boundaries of the integration range \citep{CC70}:
\begin{equation}
\Bigg[\phi\frac{\partial\eta}{\partial x} + \eta(1+
 \eta)\Bigg]_{x=x_{min},x_{max}} = 0\,.
\label{eq_bndry}
\end{equation}    
This choice of boundary conditions formally satisfies 
the requirement of the problem when 
we integrate the Kompaneets equation in the case of Bose-Einstein 
like distortions (with a frequency dependent chemical potential, 
$\mu=\mu(x)$, vanishing at very low frequencies). In fact, such distorted 
spectra are indistinguishable 
from a blackbody spectrum at sufficiently high and at low 
frequencies.

Of course, it is possible to make a different choice of the boundary 
conditions by selecting  Dirichlet like conditions. 
In this case the photon occupation number 
at the boundaries of the integration interval does not change 
for the whole integration time.
(In general cases, keeping constant conditions at the boundaries 
could be dangerous for the continuity of the solution. Nevertheless,
for some specific problems this condition can work -- typically
for problems with constant $\phi$).

\section{A detailed view on KYPRIX}
\label{sec:detailed}

The code KYPRIX has been written to solve the Kompaneets equation in 
many kinds of situations. 
The physical processes that can be 
considered in KYPRIX are: Compton scattering, bremsstrahlung, radiative 
Compton scattering, sources of photons, energy injections without photon 
production, energy exchanges (heating or cooling processes) associated to 
$\phi \ne 1$ at low redshifts, radiative decays of massive particles, 
and so on (see e.g. \cite{DB93} for some applications).
This code could be easily implemented 
to consider other kinds of physical processes.
Various kinds of initial conditions for the problem can be considered
and many of them have been already implemented in
KYPRIX. The first obvious case is a pure Planckian spectrum. 
Several ways to model an instantaneuos heating implying 
deviations from the Planckian spectrum have been introduced: 
a pure Bose-Einstein (BE) spectrum or a BE spectrum modified to 
become Planckian at low frequencies 
(this option could be exploited to integrate the 
Kompaneets equation with a constant $\phi$ and constant boundary conditions); 
a grey-body spectrum; a superposition of blackbodies.

The data are saved into five files. \\
DATI. This file contains the information about the specific parameters of 
the considered problem with a general description of its main aspects.\\ 
DATIP. In this file we give the evolution of interesting quantities, like 
time, redshift, $\phi$, 
and another quantities inherent to physical and numerical aspects of 
the problem (see also Sect.~\ref{sec:nrgy_test}).\\
DATIG. It contains: the grid of points for the $X$ axis used by 
the main program (remember that we are using a dimensionless 
frequency), a Planckian spectrum at temperature $T_0$ and the solution 
vector $U$ (that is to say $\textrm{log}(\eta)$) 
at $y = 0$ (starting time).\\
DATIDE. This is the fundamental output file: 
it gives the solution of the Kompaneets equations at the desired cosmic 
epochs.\\
DATIT. It is similar to the file DATIDE, but it contains the solution 
in term of brightness temperature (i.e. equivalent thermodynamic 
temperature; see Eq.~(\ref{eq_Ttherm}) in Sect.~\ref{sec:output}).\\

\subsection{Main subdivisions}
\label{sec:subdivisions}

The code is divided in several sections and, from a general point of view, 
is structured as described here below.\\
1. Main program, in which many actions can be carried out: 
choice of the physical processes, choice of the 
cosmological parameters, initial conditions,
characteristics of the numerical integration 
(accuracy, number of points of the grid), 
time interval of interest, 
choice of the boundary conditions,
chemical abundances, ionization history.\\
2. Subroutine PDEDEF. It is the subprogram where the problem is 
numerically defined. This subroutine is also 
divided in subsections to allow modifications in a simple and practical 
way.\\
3. Subroutine BNDARY. Here the boundary conditions are numerically
specified.\\
4. Subroutines and auxiliary functions to perform specific 
calculations.\\

\subsection{Technical specifications and code implementation}  
\label{sec:tec_spec}

The first version$^3$\footnote{$^3$Written in 1989 by C. Burigana.} 
of  KYPRIX worked with the Mark 8 version of the NAG 
numerical library and were based on the routine 
D03PGF. The version of the NAG
numerical library currently distributed is the 
Mark 21.
Therefore an update of the code KYPRIX  
is necessary to adapt it to this new package.

When KYPRIX starts running it asks all the input data: 
from the declarations of the output files' names to 
the integration accuracy and features. In the following subsections we give a 
description of the various aspects of the code (and of its update),
trying to give relevant hints about computational aspect of the code.

\subsubsection{Grid}
\label{sec:grid}

The frequency integration interval is divided in a grid of points (the 
mesh points): larger the number of points smaller the adopted 
frequency step. We adopted an equispaced grid in $X$.\\
It is possible to used a very dense grid 
(for example 36001 mesh points corresponding to 36000 frequency 
steps).
In general, it is necessary to use at least 3001 mesh points 
to have a solution accurate enough.

We found an important difference between the two NAG versions, 
not reported in the documentation of the routine 
D03PCF. In the first version (D03PGF), the subroutine where the 
partial differential equation is defined adopted 
the same mesh points defined in the main program. 
In the Mark 21 version the calculation is carried out in a different 
manner: 
the mesh points used in the subroutine PDEDEF is shifted of half 
spatial step with respect to the mesh defined 
in the main program. In this way, the mesh points in the PDEDEF 
subroutine will be exactly in the middle of the steps 
defined in the grid of the main program. For this reason, 
the limit of the integration interval are not considered 
in the mesh points in the subroutine PDEDEF and 
they are used only for the boundary conditions.\\
The effect of this feature implies the definition of new 
parameters that play a fundamental role in 
the subroutine PDEDEF. 
The integral quantities in the Kompaneets equation 
(necessary to define the radiative Compton 
term in the kinetic equations and the electron temperature) are computed 
once for any time step, inside the PDEDEF subroutine. 
For this computation,
arrays of dimension equal to the number of mesh points  
of the $x$ variable as defined in the PDEDEF subroutine
 are used.
Therefore, a particular care must be taken in the definition 
of the dimension of the 
arrays defined in KYPRIX. 
Those used 
in the main program have dimension equal to the number of points of 
the mesh defined in the main program. The same 
dimension is given for the arrays defined for the boundary conditions. 
On the other hand, the major number of arrays 
are used in the PDEDEF subroutine to compute the integral 
quantities. The 
``inner'' grid adopted in the PDEDEF subroutine is based on mesh points in 
the middle of the spatial 
steps of the main program grid, so the two grids can not 
work with the same point number; in fact, the arrays used in the 
PDEDEF subroutine have dimension $NPTS - 1$. 
Therefore, in the main program and in the subroutine BNDARY we have  
to work with arrays based on the formula:
\begin{equation}
X(I) = A + (I - 1) \times \frac{(B - A)}{(NPTS - 1)}\,,
\label{eq_grid_old}
\end{equation}
$$
\textrm{with}\, 1\le X\le NPTS\,,
$$
to define the correspondence between the grid of 
$NPTS$ points and the $X$ position, while we need 
another expression able to shift of half 
step the grid in the PDEDEF subroutine and based on 
$NPTS - 1$ mesh points:
\begin{equation}
X(I) = \Bigg(A + (I - 1) \times \frac{(B - A)}{(NPTS - 1)}\Bigg) + 
\Bigg(\frac{(B - A)}{2(NPTS - 1)}\Bigg)\,,
\label{eq_grid_new}
\end{equation}
$$
\textrm{with} \; \; 1\le X \le NPTS -1.
$$
For continuity reasons, we need
to define (according to the choices made in the main program) 
the solution vector, 
containing the photon initial distribution function, 
at the beginning of the integration also according to this grid 
definition. 
This vector is used by the PDEDEF subroutine
as initial spectrum adopted for the computation of the 
rates of the physical processes and, of course, it is then renewed 
at every time step incrementation.\\

\subsubsection{Output}
\label{sec:output}

Concerning the output files, the update version of KYPRIX stores a 
new vector containing the ``inner'' $X$ 
grid used by the PDEDEF subroutine, 
XXGR (XGR refers to the main program $X$ grid).

In addition, we preferred to have the 
possibility to perform the conversion of the solution 
into equivalent thermodynamic temperature directly 
into the code and save it in a new output file (DATIT). 
The conversion relation is:

\begin{equation}
T_{term,equiv} = \frac{x T_0}{\textrm{ln}(1 + 1/\eta)}\,
\label{eq_Ttherm}
\end{equation}
(we remember that in the code $X = \textrm{log}_{10}(x)$ 
and $U = \textrm{log}_{10}(\eta)$). 

\noindent
The fundamental reason 
to perform this conversion directly in the code is associated
to the extreme
accuracy required for the solution in the case 
of very small distortions, of particular interest given the
FIRAS results \citep{fixsen96}.
During the first tests, the conversion of the solution in brightness 
temperature was performed at the same time of the solution visualization, 
through the IDL visualization program. 
The saving of the solution into files is typically performed 
not for all the points of the grid 
but for a reduced grid 
of, for example, 300 equidistant points along the original grid
to avoid to store files of large size, 
useless for our scope, given the interest 
for the CMB continuous spectrum
(by definition, the Kompaneets equation is not 
appropriate to treat recombination lines).
If the considered distortions were very small then the solution 
at each specific ``inner'' grid point could be affected by a 
numerical uncertainty not negligible in comparison with the 
very small deviations from a Planckian spectrum relevant in this cases.
This numerical error is greatly reduced (becoming negligible for our 
purposes) by the averaging over a suitable number of grid points.
Of course, the storing of the solution directly on a limited number 
of grid points makes this averaging no longer possible on the stored data.
It were then necessary to average the 
solution values in intervals corresponding to the output $x$ grid
directly into the code. 
Anyway, in many circumstances the diagram shape derived applying the
conversion to brightness temperature only on the stored averaged solution
still deviates at high frequencies
from the effectively computed solution displayed
by considering all the 
``inner'' grid points because of the high gradients in the 
photon distribution function and/or in the brightness temperature
that makes difficult, or impossible, to find a general rule for the 
solution binning that simultaneously works properly for the two 
solution representations.
This problem is avoided converting the solution vector 
in equivalent thermodynamic temperature before of the binning of its 
values and then applying the binning to the equivalent thermodynamic 
temperature. 
The result is then a brightness temperature 
diagram very clean and precise, even for very small distortions.

Other minor changes are made about the output data, where we passed from 
real to double precision, and for the saving frequency into the output files.

\subsubsection{Equation formalism}
\label{sec:formalism}

A necessary update of the code has been performed to adapt it 
to the different formalism adopted by the new version of the NAG routine.
This regards the expression of the Kompaneets equation in the PDEDEF 
subroutine. 
In particular, the D03PGF routine adopted the following expression 
of the partial differential equation:
\begin{equation}
C_i \frac{\partial U_i}{\partial Y} = 
X^{-m} \sum_{j=1}^{NPDE} \frac{\partial}{\partial X}\Bigg[ X^m G_{ij} 
\frac{\partial U_j}{\partial X} \Bigg] + F_i\,,
\label{eq_d03pgf}
\end{equation}
where $i = 1,2,...,NPDE$ (number of partial differential equations); $C_i, 
F_i$ depends on $X,Y,U,\partial U/\partial X$; 
$G_{i,j}$ depends on $X,Y,U$ and $U$ is the set of solutions values 
$(U_1,U_2,...,U_{NPDE})$. 

The expression now adopted by the D03PCF routine is represented 
by Eq.~(\ref{pdedef}).

It is simple to translate the code from the old to the new formalism. 
In the considered case $NPDE = 1$. 
In this case, we have simply that 
$R_1$ contains both the function $G_1$ and 
the vector solution derivative with respect to $X$ according to:

\begin{equation}
R_1 = G_{11} \times \frac{\partial U_1}{\partial X}\,.
\label{eq_r1}
\end{equation}

At this point, it is necessary to apply only 
the following substitutions:

\begin{equation}
Q_1 = - F_1\qquad\textrm{and}\qquad P_{11} = C_1\,. 
\label{eq_q1}
\end{equation}

With respect to this formalism, it is not difficult 
to adapt the various terms of the Kompaneets equation to 
the D03PCF routine. The terms that describe Radiative Compton, 
bremsstrahlung, (optional) electromagnetic processes and part 
of the contribution of the Compton scattering are counted in the 
function $Q_1$. Instead, the second derivative of the solution 
vector with respect to $X$, which represents part of the inverse Compton rate, is 
counted in the function $R_1$. 
So, according to these settings, we can write:
\begin{equation}
Q_1 = FC + FBREM + FRAD + FDEC\;,
\label{eq_quno}
\end{equation}
where $FC$ stands for the contribution of the Compton scattering, 
$FBREM$ the bremsstrahlung one, $FRAD$ the radiative Compton 
one and $FDEC$ represents the contribution of radiative decaying of particles. 
These terms are written in the form:
$$
FC = \bigg[ \phi \frac{\partial U}{\partial X} \left( \frac{\partial U}
{\partial X} + 3 \right)
+ 10^X \left(\frac{\partial U}{\partial X} + 4 \right)
$$
\begin{center}
\begin{equation}
\qquad\qquad\qquad+ 2 \times 10^X 10^U \left(\frac{\partial U}{\partial X}
 + 2 \right)\bigg] \frac{1}{{\rm ln}(10)}
\end{equation}
\end{center}
and
$$
FBREM+FRAD= \frac{\left({e^{10^X}}\right)^\phi}{10^{3X}} \times
\left\{\frac{1}{10^U}-\left[{\left(e^{10^X}\right)}^{{\phi}^{-1}} - 1\right]\right\}
$$
\begin{equation}
\times \left(FF0 \times W \times 1.5 \phi^{-1/5} \times FGAUNT +
\frac{DC0}{W}I_1 \times GDC \right)\,,
\end{equation}
where $FF0$ and $DC0$ are the coefficients for the rates of 
bremsstrahlung and radiative Compton respectively; 
$FGAUNT$ and $GDC$ represent the Gaunt factor corrections for 
bremsstrahlung and radiative Compton, respectively; $I_1$ is an 
integral quantity refreshed at every time step 
(see also Sect.~\ref{sec:rad_comp}).

The inverse Compton contributes to $R_1$ in this way:
\begin{equation}
R_1 = \frac{\partial U}{\partial X} \times \phi \times {{\rm ln}(10)}^{-2}\;.
\end{equation}

Finally, we can put $P_{11} = 1$.

Furthermore, it is possible to add other source 
terms$^4$\footnote{$^4$For example, an already implemented subroutine
models a term, called here $FDEC$, to be added in Eq.~(\ref{eq_quno}) 
which acconts for contributions of possible radiative decays of 
massive particles in the primordial Universe.}
in Eq.~(\ref{eq_quno}).

\subsubsection{Boundary conditions}
\label{sec:bndry_form}

Also notable are the differences between the input 
expressions defining the boundary conditions. 
The D03PGF routine 
adopted an expression of the form:
\begin{equation}
P_i(Y) U_i + Q_i(Y) \frac{\partial U_i}{\partial X} = R_i(Y,U) \, ,
\label{eq_bndry_form}
\end{equation}
where $i = 1,2,...,NPDE$ and $P_i(Y),R_i(Y,U),Q_i(Y)$ are functions 
to be defined. A quite different notation is 
used to provide 
the boundary conditions in the D03PCF routine: 
\begin{equation}
\beta_i(X,Y) R_i(X,Y,U,U_X) = \gamma_i(X,Y,U,U_X) \, ,
\label{eq_beta}
\end{equation}
where $i = 1,2,...,NPDE$ and $\beta_i(X,Y) R_i(X,Y,U,U_X)$ 
and $\gamma_i(X,Y,U,U_X)$ are functions to be defined 
$(U_X \equiv \partial U/\partial X)$.\\
As a consequence of this notation, Neumann like 
boundary conditions can be now specified according to the 
expression:
\begin{equation}
\beta(1) = 1
\end{equation}
\begin{equation}
\gamma(1) = - XVA \times (10^{U(1)} + 1) \times \mbox{ln}10^{-2}
\label{eq_gamma}
\end{equation}
where $XVA$ is the vector related to the $X$ position computed in 
A and the dimension of both 
the equations corresponds to the differential 
equation number. 
Similar conditions are defined for the other extreme 
of the integration interval [A,B].\\

\subsubsection{Accuracy parameters}
\label{sec:acc}

Another considerable difference between the two library versions 
regards the definition of 
the integration accuracy parameter. 
D03PGF used three parameters for monitoring the local error estimate
in the time direction, supplying a good versatility. 
RELERR and ABSERR were respectively the quantity for the 
relative and absolute component to be used in the error test. 
The third parameter, INORM, was used to 
define the error test. 
If E$(i,j)$ is the estimated error for $U_i$ (the vector solution) 
at the $j-th$ 
point of the $X$ grid, then the error test was:
\begin{itemize}
\item INORM $= 0 \Rightarrow \mid\textrm{E}(i,j)\mid \le 
\textrm{ABSERR} + \textrm{RELERR} \times \mid U(i,j) \mid$
\item INORM $= 1\Rightarrow \mid\textrm{E}(i,j)\mid \le 
\textrm{ABSERR} + \textrm{RELERR} \times {\textrm{max}_y} 
\mid U(i,j) \mid$
\item INORM $= 2 \Rightarrow \|\textrm{E}(i,j)\| \le 
\textrm{ABSERR} + \textrm{RELERR} \times \| U(i,j) \|$.
\end{itemize}
Instead, according to the new library version we have to define only one 
parameter ACC, a positive quantity that monitors 
the local error in the time integration. 
If E$(i,j)$ is defined as above, then the error test is:
\begin{equation}
\mid\textrm{E}(i,j)\mid = \textrm{ACC} \times (1 + \mid U(i,j)\mid) \, .
\label{eq_err}
\end{equation}

\noindent
Note that this is equivalent to the error test implemented 
in the D03PGF routine in the case INORM~=~0 and
$\textrm{ABSERR} = \textrm{RELERR}$ $(=\textrm{ACC})$.

\subsubsection{Electron temperature}
\label{sub:te}

During the numerical integration, 
some subprograms use the 
distribution function calculated at that time to compute $\phi$. 
The integrals to be computed are those that we 
find in the expression for $\phi_{eq,C}$: 
\begin{equation}
\phi_{eq,C} = \frac{T_e}{T_{\gamma}} = 
\frac{\int_0^{\infty}\eta(\eta + 1)x^4\textrm{d}x}{4\int_0^{\infty}
\eta x^3\textrm{d}x}\,.
\label{eq_phi_eq}
\end{equation}
In this calculation, the integration range is obviously the integration 
interval considered for the problem: 
$A \leq X \leq B$ (that, in terms of mesh ordering, corresponds to the 
range between 1 and $NPTS$ or $NPTS-1$). 
For computing 
these integrals,  all the points 
of the grid are used. The integration is based on 
the NAG D01GAF routine, suitable for tabulated functions. 
The update value of $\phi$ 
is also used in the boundary conditions.

In the previous version of the code KYPRIX, the computation of 
integral quantities were performed through a specific modification of the 
NAG package implemented by the KYPRIX code author that allowed 
to recover the whole vector solution at each time step 
in the subroutines (and in particular in PDEDEF), while the 
original package made only available in PDEDEF the solution 
separately at each grid point (being in fact the package 
originally designed for ``pure'' partial differential equation,
without terms involving integrals of the solution). This modification,
possible thanks to the availability of the NAG sources (and, in practice,
thanks to the relative simplicity of the early library versions),
permitted to update the integral quantities perfectly according
to the ``implicit'' scheme adopted by the code for the integration in 
time.
This is no longer feasible. Therefore, the update of the integral
quantities must be now performed with a ``backward'' scheme, saving
the solution at the previous time step in a proper vector
and using it in the computation at the given time step.
As well known, ``backward'' schemes are typically less stable
than implicit schemes. 
And, in fact, we verified in some cases   
the difficulty of the 
D03PCF routine to work implementing the update of the quantities 
corresponding to the integral 
terms in the Kompaneets equation (and in particular of $\phi$)
for each time step. 
This were likely due to numerical instabilities.

We have then introduced a new integer control parameter into the code: 
STEPFI.
It determines the frequency for the update the dimensionless electron 
temperature $\phi$, 
relevant, of course, in the case we want to perform an integration with a 
variable $\phi$.
We have checked that updating the integral
terms in the Kompaneets equation not at every time step, but after 
a suitable number of time steps does not affect the accuracy of the 
solution. This is due to the fact that the time increasing in the code
is performed with very small steps while the physical variation
of $\phi$ occurs on longer timescales$^5$\footnote{$^5$Of course, for 
physical processes with a stronger variation of the electron temperature, the 
accuracy parameter (see previous subsection) should be good enough
to force the code to adopt sufficiently small time steps.}.

See Sect.~\ref{sec:test_phi} for tests regarding the 
implications of this new implementation of the electron temperature 
evolution.

\subsubsection{Radiative Compton}
\label{sec:rad_comp}

In the computation of the radiative Compton term there is an integral term, 
$I_1$, given by the numerator of the right member of Eq.~\ref{eq_phi_eq}
(see Eq.~(16) in \cite{BDD95}),
so it is necessary to harmonize its update
according to the parameter STEPFI discussed in the previous subsection. 
In fact, 
a possible asynchronous update of it and 
$\phi$ could create numerical instabilities 
and the crash of the code run, as physically evident from the
great relevance of both radiative Compton term 
(at least at high redshifts)
and electron
temperature for the evolution of the low frequency region of the 
spectrum.\\

\subsubsection{Integration routines}
\label{sec:routines}

The global accuracy of the code KYPRIX depends on the accuracy 
of the solver for the partial differential equation as well as on the 
accuracy of all the other routines dedicated to different specific
computations.
In this section we focus on the routines of numerical integration 
used in the KYPRIX.
As discussed in the previous subsections the D01GAF routine has been 
used for tabulated functions. On the other hand, 
KYPRIX involves the computation of integrals
of various functions defined by analytical expressions, namely
the parameters characterizing different types of initial conditions for 
the distorted spectra, as the amount of fractional injected energy 
and the electron temperature, and the relationship between the different 
time variables entering in the code (see also 
Sect.~\ref{sec:cosm_const}).
Ultimately, the better accuracy in these specific computations
implies a better global accuray of KYPRIX.
In the early release of the code KYPRIX 
the NAG D01BDF routine was used to calculate
integrals of a function over a finite interval. 
The same task can be carried out by the D01AJF routine. 
This code offers a better 
accuracy than D01BDF (D01AJF is in fact suitable also 
to integrate functions with singularities, 
both algebraic and logarithmic). After the routines 
substitution, the results showed a great increasing of accuracy.
In particular, this improvement offers the possibility to investigate
also very small distortions that requires a very precise determination
of all the relevant quantities because the absolute numerical error
of the integration must be much smaller than the (very small
quantities) of interest in these cases. 
In particular, the quantity 
$\Delta \epsilon_r/\epsilon_i$ (where $\epsilon_r$ is the actual 
density energy and $\epsilon_i$ is 
the energy density corresponding to the unperturbed distribution function 
just before the energy injection)
must be constant during all the 
integration process in the absence of energy injection terms,
according to the energy conservation.
The precision increase on the computation of this quantity was 
noteworthy, keeping now always inside a few percent of the physical 
value (and its possible physical variation; see 
Sect.~\ref{sec:nrgy_test}) 
of the same quantity independently of the magnitude of the 
considered distortion, allowing at the same time to accurately 
check the global accuracy of the code, improved thanks 
to the better computation of all the 
integral terms appearing in the Kompaneets equation.
The remarkable improving in energy conservation 
even in the case of very small distortions 
is an important feature of the new version of KYPRIX
that makes is applicable to a wider set of cases.\\

\section{New physical options}
\label{sec:new-opt}

\subsection{Cosmological constant}
\label{sec:cosm_const}

About ten years ago, the relevance of the cosmological constant term (or 
of dark energy contribution) has been probed by a wide set of astronomical 
observations of type Ia Supernovae \citep{SN1,SN2}.
We have then updated the numerical integration code KYPRIX 
to include the cosmological constant in the terms controlling the
general expansion of the Universe.
In particular the input background cosmological 
parameters considered in the code are now: $T_0, \kappa,  
h [=H_0/(100{\rm Km/s/Mpc)}], 
\Omega_{nr}, \Omega_b, \Omega_{\Lambda}, \Omega_K$,
i.e. the present CMB temperature, the contribution of massless 
neutrinos, the Hubble constant, the (non relativistic) matter 
and baryon energy density, the energy densities corresponding 
to cosmological constant and curvature terms.

In order to compute the proper cosmic evolution of the various terms,
an (increasing with time)  
scale factor parameter $\omega$ \citep{SS83}, 
defined by
\begin{equation}
\omega = \frac{a}{a_1} \equiv \frac{m_e c^2}{k T_0}\frac{1}{1 + z} = 1.98 
\times 10^9 \Theta (1 + z)^{-1},
\label{eq_omega}
\end{equation}
has been adopted (here $\Theta \equiv {T_0/3}^{\circ}K$ and the index 1 
is referred to a particular epoch, when the CMB energy 
density was equal to the electron 
mass$^6$\footnote{$^6$So, the parameter $\omega$ is analogous to 
the scale factor $a$, but normalized at the epoch in which $a=a_1$, that 
is to say when $kT=m_ec^2$.}: $k_T(a_1) = m_ec^2$). 
To write a suitable expression for its time evolution we have to introduce 
two new key parameters
\begin{equation}
\beta = \frac{\rho_{m1}}{\rho_{r1}} = 
3.5 \times 10^{-6} \frac{h^2}{\Theta^3}\;\Omega_{tot}
\label{eq_beta_nrgy}
\end{equation}
that is the initial ratio between matter energy density 
and radiation energy density, and
\begin{equation}
\frac{1}{\tau_{g1}} = \bigg(\frac{8\pi}{3} G \rho_{r1}\bigg)^{1/2} = 
\bigg[\frac{8\pi}{3}G\frac{a}{c^2}\bigg(\frac{m_ec^2}{k}\bigg)^4\bigg]^{1/2} 
= 0.076\textrm{s}^{-1}
\label{eq_taug}
\end{equation}
defined as a initial gravitational time scale. The quantities with the 
index 1 refer to the epoch when 
$a = a_1$, with the index 0 when $t = t_0$ (today); 
$\rho_{r1}$ and $\rho_{m1}$ are related 
to $\rho_r$ and $\rho_m$ by  
\begin{equation}
\rho_r = \rho_{0r}\bigg(\frac{\omega_o}{\omega}\bigg)^4 =
 \rho_{r1}\frac{1}{\omega^4};\;\;\;
\rho_m = \rho_{0m}\bigg(\frac{\omega_0}{\omega}\bigg)^3 = 
\rho_{m1}\frac{1}{\omega^3} \, ,
\label{eq_rho}
\end{equation}
respectively.

Now we can define an equation for the evolution of $\omega$:
\begin{equation}
\frac{\dot{\omega}}{\omega} = \bigg[\frac{8\pi}{3} G \rho(\omega)\bigg]^{1/2} =
\label{eq_evol_omega}
\end{equation}
$$
\;\;\;\;\frac{8\pi}{3}G\bigg[\frac{\rho_{r1}\kappa}{\omega^4} + 
\frac{\rho_{m1}}{\omega^3} + 
\frac{\rho_{K1}}{\omega^2} + \rho_{\Lambda}\bigg],
$$
where we have included the contribution of massless relativistic neutrinos 
in the term $\kappa$ (see also footnote 1; the term $\kappa$ should be 
properly evaluated considering also possible energy injections after
neutrino decoupling).

After some calculations, we can finally write the 
completed and updated expression of 
${\textrm{d}t}/{\textrm{d}\omega}$: 
\begin{equation}
\frac{1}{\dot{\omega}} = \frac{\tau_{g1}\;\omega}{\bigg[1 + \beta\;\omega\bigg(1
 + \frac{\Omega_{K/m}\;\omega}
{2.164 \times 10^9} + \frac{\Omega_{\Lambda/m}\;\omega^3}{2.164 \times
 10^{27}}\bigg)\bigg]^{1/2}}\;,
\label{eq_omega_finale}
\end{equation}
where $\Omega_{x/y} = {\Omega_x}/{\Omega_y}$.

The equation for $\dot{\omega}$ has to be inserted in the expression 
giving $\textrm{d}y=a_c\textrm{d}t$, 
which is inside the integral used to compute the time variable 
$y(\omega)=\int_{\omega_{start}}^{\omega}\textrm{d}y$ because we set
$y=0$ when the integration starts at $\omega=\omega_{start}$
(or equivalently at $z=z_{start}$). 
Finally, the 
expression for the time evolution of $\omega$ and $y$ 
are related by the variable change:
\begin{equation}
\textrm{d}y = a_c \textrm{d}t = a_c\;\frac{\textrm{d}t}{\textrm{d}\omega}\;
\textrm{d}\omega = 
a_c \;\frac{\omega}{\dot{\omega}}\;\frac{1}{\omega}\;\textrm{d}\omega\;,
\label{eq_dy}
\end{equation}
where $a_c = \phi/(\tau_{c1}\omega^4)$ 
($\tau_{c1} = 2.638 \times 10^{-9}\,\Theta^3/(h^2\Omega_b$)).

The implementation of the cosmological constant and curvature 
terms makes the code KYPRIX suitable to be applied to interesting 
cases at late ages, including cosmic epochs at 
$z < 1$ when $\Lambda$ supplies the greatest contribution to the expansion 
rate of the Universe.
Remarkable examples are spectral distortions
associated to the reionization of the Universe that, in 
typical astrophysical scenarios, starts 
at relatively low redshifts ($z \lsim 10-20$).
For sake of completeness, a few words about the computation of the time 
evolution in the code. The subroutine called WDIY0 is the core of the 
time evolution in KYPRIX: it computes the value of $\omega$ given a value 
of $y$. This process takes advantages of the definition of $y$ as integral of 
d$y$ and, of course, it happens at each time step. To do this, we make use 
of a double precision version of the function ZBRENT, from 
``Numerical Recipes'' \citep{press}.

\subsection{Chemical abundances}
\label{sec:chem_ab}

In this new version of the code it is possible to choose the primordial 
abundances of $H$ and $He$. We have consistently computed
the electron number density,
$n_e$, involved in the different physical processes. 
In the previous implementation it was assumed a fixed abundance of $H$ 
and $He$ (and full ionization). 
The electron number density was then given by 
\begin{equation}
n_e^{free} = n_e^{tot} \simeq \frac{\rho_b}{m_b} \frac{7}{8}\;,
\end{equation}
where $\rho_b$ is the baryon density and $m_b$ the mean mass of a baryon.
Now, denoting with $f_H$ the fraction in mass of primordial $H$ and 
considering that $n_e=n_H + 2 n_{He}$,
we have 
\begin{equation}
n_e^{free} = n_e^{tot} = \frac{1+f_H}{2}\frac{\rho_b}{m_b}\,.
\end{equation}
This obviously impacts 
the physical processes involved in the code. Note that Compton 
scattering, radiative Compton, and bremsstrahlung depend
linearly$^7$\footnote{$^7$For 
bremsstrahlung, the dependence on the densities of nuclei
is now explicit.} on $n_e^{free}$. 

\subsection{Ionization history}
\label{sec:reion}

In the last years, CMB observations are getting a very high 
accuracy, in particular regarding the angular power spectrum 
of temperature anisotropies albeit also for that of
E mode polarization and cross-correlation (see e.g. \cite{wmap5yr_nolta}). 
A detailed understanding of the cosmological reionization process 
is crucial for precisely modeling the power spectrum of CMB 
anisotropies in comparison with current data, while a better 
treatment of recombination is of great relevance in the view
of the data expected by the forthcoming ESA {\it Planck} 
satellite$^8$\footnote{$^8$http://www.rssd.esa.int/planck} 
\citep{planckbluebook}.

An accurate modeling of the ionization history is crucial also for 
the precise computation of CMB spectral distortions  
in the view of the comparison with data from 
a future generation of high sensitivity experiments.
We have then included these aspects in the current
implementation of KYPRIX. 
In particular,  
the fraction of each state of ionization of the relevant 
elements (hydrogen, $H$, and helium, $He$) 
has been implemented.

The effects of this implementation is negligible$^9$\footnote{$^9$For 
example, 
for $z \lsim 10^{-4}$ and $\Delta\epsilon_r/\epsilon_i = 10^{-5}$ 
we find that, for both BE and Comptonization like distortions, 
the rate of radiative Compton is less than 1/1000 of the rate of 
bremsstrahlung at each frequency of the grid.} in practice for the 
radiative Compton, because during the epochs this process is 
active the medium is fully ionized.
Instead, Compton scattering and bremsstrahlung rates are significantly 
influenced by this implementation.

Once introduced the electron ionization fraction in the code, $\chi_e$, 
which gives the number of electrons that takes 
part in the physical processes, we can choose different ways by which 
the active fractions of elements can play their roles in the phenomena.

Given $\chi_e$, from the charge conservation law we have a constraint 
on the number of the free ions in the considered plasma. 
The simplest way to take count of them in 
the code is to assume an equal fraction of ionization for $H$ and $He$. 
Of course, this is a toy model, but this parametrization was 
very useful to test the code.

A somewhat more accurate treatment of the physics of 
reionization/recombination processes 
implemented in the code is based on the Saha equation:
\begin{equation}
\frac{n_{i+1} n_e}{n_i} = \frac{2}{\Lambda^3} \frac{g_{i+1}}{g_i}
e^{-\frac{\epsilon_{i+1}-\epsilon_i}{k_BT}}\,,
\end{equation}
where $n_i$ is the density of atoms in the $i$-th state of ionization, 
$n_e$ is the electron density, $g_i$ is the degeneracy of states for 
the $i$-ions, $\epsilon_i$ is the energy required to remove $i$ 
electrons from a neutral atom and $\Lambda$ is the thermal 
de Broglie wavelength of an electron, defined by
\begin{equation}
\Lambda=\sqrt{\frac{h^2}{2 \pi m_e k_B T}}\,.
\end{equation}
Providing the electron ionization fraction defined 
as$^{10}$\footnote{$^{10}$It is common to find in the literature 
$\chi_e$ normalized to the $H$ number density. 
Obviously, the code can switch between the two conventions.}
\begin{equation}
\chi_e \equiv \frac{n_e^{free}}{n_e^{tot}}\;, 
\end{equation}
we can compute the unknowns  
$\chi_H, \chi_{H^+}, \chi_{He}, \chi_{{He}^+}, \chi_{{He}^{++}}$,
defined as the 
relative abundances of the different ionization states of each 
element with respect to the global number density of the element. 
The Saha equation provides the ratio between two state of ionization 
of a single specie, once given the electron density and the temperature:
\begin{equation}
\frac{n_{H^+}}{n_H}\;\;,\;\;\frac{n_{{He}^+}}{n_{He}}\;\;,\;\;
\frac{n_{{He}^{++}}}{n_{{He}^+}}\;\;.
\label{ratios}
\end{equation}

To recover all the unknowns we need other relations. 
These additional conditions are provided by 
the charge conservation and the nuclei conservation. 

From charge conservation it is possible to recover the contribution of 
the electrons related to a single specie to the total number of free electrons. 
The latter is of course related to the ionization fraction 
of all the involved species 
and can be written as: 
\begin{equation}
n_e^{free} = \frac{\rho_b}{m_b} \left[ \chi_{H^+} f_H + 2\left(\frac{1 - f_H}{4}\right) 
\left(\chi_{{He}^+} + \chi_{{He}^{++}}  \right) \right]\;,
\label{ne_free}
\end{equation}
where $f_H$ is the fraction of primordial $H$ described in the previous section. 
In Eq.~(\ref{ne_free}) it is possible to identify the number of 
electrons coming from $H$ and from $He$. 

The nuclei conservation law separately for $H$ and $He$
is expressed by
\begin{equation}
n_H^{tot} = \frac{\rho_b}{m_b} \left[ f_H (\chi_H + \chi_{H^+}) \right]\;,
\end{equation}
\begin{equation}
n_{He}^{tot} = \frac{\rho_b}{m_b} \left[\left(\frac{1 - f_H}{4}\right) (\chi_{He} + 
\chi_{{He}^+} + \chi_{{He}^{++}}) \right]\;.
\end{equation}

Providing the electron ionization fraction and the temperature, it is 
possible to build two separate systems 
(one for $H$ and one for $He$) in order to recover the 
considered ionization fractions.

The code can ingest a table with the desired evolution of $\chi_e$
and, as necessary for a physical modeling of cosmological reionization,
of the electron temperature. 

Obviously, we have implemented the best way to perform the exact 
calculation of the rates of considered processes in scenarios involving 
reionization/recombination which is that of 
using a co-running code, coupled to KYPRIX, able to supply the 
ionization fraction for all the species.
For the recombination process, 
we developed an interface that allows the code to call an external program, 
in our case RECFAST \citep{recfast1}, and run it with the same 
cosmological parameters selected for KYPRIX. 
Then KYPRIX will read the output table from RECFAST and will use 
the ionization fractions for 
the calculation of the rates of the processes, allowing a more detailed 
estimation of the spectral distortion arising during the 
recombination process \citep{recfast2}. If we are interested in
standard recombination, we adopt the equilibrium electron 
temperature (Eq.~\ref{eq_phi_eq}).

A consistent modeling of a modified recombination should provide
the evolution of both electron temperature and ionization fractions
(at least $\chi_e$).

\section{Tests on code performance, reliability, and accuracy}
\label{sec:test}

Once terminated the updating of the numeric integration code, 
we have carried out many accuracy and performance tests. 
Obviously, we have checked 
the physical meaning of the numerical solutions provided by the code
comparing them
with the existing analytical solutions that 
in some cases can be considered as good approximations
of exact solutions.
In general, a code of good quality must have
high numerical precision compared to the knowledge, 
both theoretical and observational,
of the considered problem.
Note also that various routines, mainly from the NAG package 
albeit also from the ``Numerical Recipes'' \citep{press} package, 
are used here for several specific computations. 
Since different routines can solve the same mathematical problems
using different numerical methods and/or implementations
with different settings and input parameters,
we have verified that the adopted routines allow the 
appropriate accuracy and efficiency. 

In the following subsections we will report on some 
specific quality tests of the numerical solutions.
We comment now briefly on the code computational time.

In order to evaluate the global CPU time of the code  
we performed many runs with very different settings.
These times are carried out testing the code on a 
machine with 4 Alpha CPUs, but effectively using only one CPU 
(now we are running the code on IBM Power5 Processors).
The code global CPU time ranges from few minutes, 
for cases in which the integration 
starts at low redshifts, to about 5 hours, for some cases starting at 
very high redshifts ($y(z) \simeq 5$). 
There are many
variables that take role in determining the global CPU time.
The complete Kompaneets equation involves in fact several terms. 
In the code KYPRIX we can select the physical processes 
to be considered in the numerical integration and the global CPU time 
increases with the number of activated processes.

Of course, the global CPU time depends on the
parameters related to the numerical integration characteristics.  
The number of points adopted for the $X$ grid 
(see Sect.~\ref{sec:grid})
has a great influence on
the global CPU time. 
Clearly,  the integration accuracy improves with $NPTS$ 
and in many cases of interest it must be set to a very large value. 
We find that the global CPU time
is approximately proportional to $NPTS$ ($t_{CPU} \propto NPTS$).

The parameter that plays the most relevant role 
in determining the global CPU time
is the accuracy required for the time integration.
The final solution precision depends on the value of the corresponding 
parameter ACC.
Only for very high accuracy 
(ACC~$\lsim 10^{-12} - 10^{-14}$) the CPU 
time reaches the duration of some hours while keeping 
ACC~$\sim 10^{-5}$ the integration is carried out in few minutes. 
Anyway, the limits imposed by CMB spectrum observations drive us to investigate 
in particular on small distortions. 
It is then necessary to work with low values of ACC (in general, 
$\la 10^{-8}$).

After the assessment of the better choice for the use of 
the parameters of the various 
numerical routines and the definition of the accuracy of the  
integration in time, 
we have carried out several tests in order to 
verify the physical validity of the code results.\\

\subsection{Energy conservation}
\label{sec:nrgy_test}

\begin{figure}
 \centering
 \includegraphics[width=1.05\hsize]{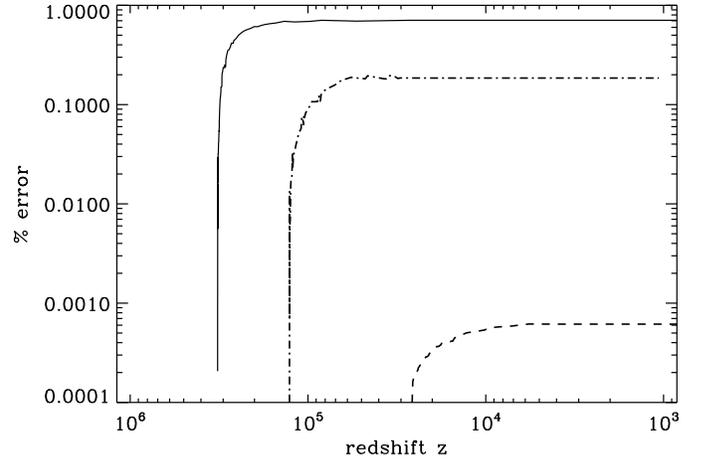}
 \caption{Error in the energy conservation expressed in terms of
relative (\%) deviation from its input value of the quantity
$\Delta\epsilon_r/\epsilon_i = 10^{-5}$. The accuracy of the 
integration in time is set to ACC$=10^{-12}$
(see also the text for further details on computation).
Solid, dot-dashed, and dashed lines refer to an energy injection 
occurred respectively at $y_h(z) \simeq 1.5, 0.25, 0.01$.
Obviously, this error further decreases 
improving the accuracy parameter adopted in the numerical integration.}
\label{fig:energy_cons}
\end{figure}

In the code output file DATIP we store values of several parameters of interest.
Two of them provide very useful information on the 
goodness of the numerical integration. 
The first one is the ratio,
${\epsilon_r}/{\epsilon_{i}}$, 
between the radiation energy density at each time and the energy density
corresponding to the unperturbed distribution function before 
the distortion$^{11}$\footnote{$^{11}$For example, for a Bose-Einstein 
distorted spectrum 
with a dimensionless chemical potential $\mu_0$
produced by an energy dissipation with negligible photon production
${\epsilon_r}/{\epsilon_{i}} = 
f(\mu_0)/\varphi^{4/3}(\mu_0) \,$ 
(see e.g. \cite{SZ70}, \cite{DD77} and
also equations (8) and (9) in \cite{BDD91a}
for the definition of $f(\mu_0)$ and $\varphi(\mu_0)$
for arbitrary values of $\mu_0$; 
for $\mu_0 \ll 1$, $f(\mu_0) \simeq 1-1.11\mu_0$
and $\varphi(\mu_0) \simeq 1-1.368\mu_0$).}
in the absence of dissipation processes, the exact energy conservation 
is represented 
by the constance of this ratio during the integration.
To quantify the accuracy supplied by KYPRIX, 
the values of $\epsilon_r/\epsilon_i$ are stored
at the starting of the integration and at many following times. 
In order to estimate the precision of the energy conservation, 
we define the quantity:
\begin{equation}
ERR_{\epsilon} = \frac{|(\epsilon_r/\epsilon_i)_{t=t_{start}} -
 (\epsilon_r/\epsilon_i)_{t>t_{start}}|}
{(\epsilon_r/\epsilon_i)_{t=t_{start}} - 1}\, ,
\label{eq_epsilon}
\end{equation}
that gives the relative induced error in the energy conservation
with respect to the
initial value of $\Delta\epsilon_r/\epsilon_i$ or, more formally, the 
relative error induced by numerical uncertainty on the amount of
fractional injected energy.
Typical results obtained starting from a superposition of blackbodies 
are reported in Fig.~\ref{fig:energy_cons}.

Since the same absolute numerical integration error 
corresponds to a larger relative error for a smaller distortion, 
i.e. for smaller $\Delta\epsilon_r/\epsilon_i$ in these models, 
we could in principle expect a relative degradation of the energy 
conservation for decreasing distortion amplitudes.
Our tests indicate in fact 
an increasing
of the relative degradation of the energy conservation 
for decreasing distortions when the same accuracy parameters are 
adopted.
On the other hand, one can select them according the specific problem.
For example, considering an energy injection of 
$\Delta\epsilon_r/\epsilon_i = 10^{-5}$ 
occurred at $z \sim few \times 10^4$, 
an accuracy equal to $ACC = 10^{-8}$ is fully satisfactory 
for a very precise calculation of the 
photon distribution function, 
while for earlier processes, as for $z \gtrsim few \times 10^5$, 
and for the same injected fractional energy, 
the accuracy parameter needs to be set to 
$ACC \leq 10^{-12}$ in order to assure a 
calculation with 
$ERR_{\epsilon}$ less than 1\%.
Anyway, we find that for suitable choices of the integration accuracy 
parameters, $ERR_{\epsilon}$ can be kept always 
below $\simeq 0.05\%$
without requiring a too large computational time. 
Finally, we note that in some circumstances
the scheme for the electron temperature evolution
in the new version of KYPRIX (backward differences), 
different from that used in the original one (implicit scheme),
could imply some small discontinuities in the evolution 
of the electron temperature 
and of $\Delta\epsilon_r/\epsilon_i$. On the other hand, 
we have verified that this effect does not affect the very good
accuracy of the solution, because of the very small amplitude of these
discontinuities (together with their localization to a very limited 
number of time steps) and of the corresponding energy 
conservation violation.

\subsection{Comparative tests}
\label{sec:comp_test}

\subsubsection{Comparing solutions}
\label{sec:solutions}

The first kind of test is simple 
comparison between the results obtained with the update
version of KYPRIX and those obtained with the original version
for the same set of input parameters.

\begin{figure}
 \centering
 \includegraphics[width=0.82\hsize,angle=270]{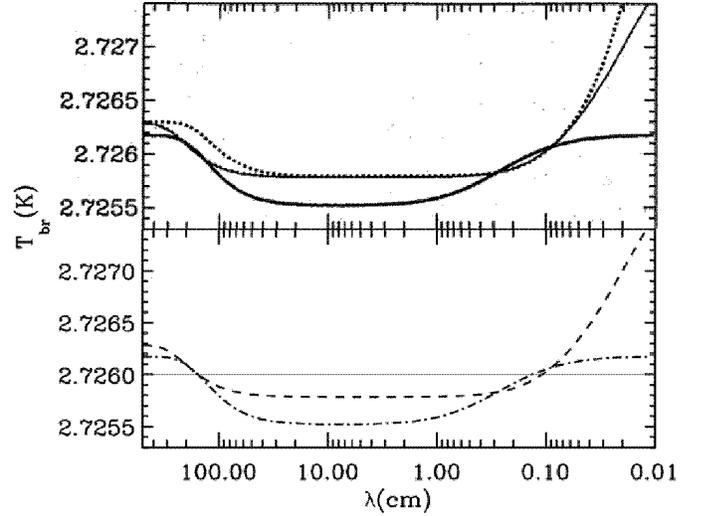}
 \caption{Comparison between the present time solution 
for the CMB spectrum obtained from the old version of 
the numerical code (top panel; 
adapted from panel a of Fig.~1 in \cite{BDD95}) 
and the current one (bottom panel).
See the text for further details on computation parameters.
Dashed (dot-dashed) line refers to an energy injection occurred at 
$y_h(z) \simeq 0.01$ ($y_h(z) \simeq 0.25$). 
Note the excellent agreement between the results of the two codes.
In the case of the 
old version of the code we report also the analytical approximation
(dotted line) described by a Comptonization spectrum 
plus a free-free distortion
(see Eq.~\ref{eq_eta_approx}).}
\label{fig:comp_sol}
\end{figure}

To this purpose, we have considered some interesting cases
carried out in the past. In particular we used the input parameters 
adopted in 
\citep{BDD95}
where also semi-analytical descriptions 
of the numerical solutions of the Kompaneets equation were reported.
We report here cases characterized by  
an amount of exchanged fractional energy
$\Delta\epsilon_r/\epsilon_i = 10^{-4}$. We started the 
integration from a redshift
corresponding to $y_h(z) \simeq 0.25$ in one case 
and to $y_h(z) \simeq 0.01$ for another. 
The input cosmological parameters are: 
$H_0 = 50,\;\widehat{\Omega}_b = 0.03,\; k = 1.68,\;T_0 = 2.726$~K. 
The results given by the update 
version of KYPRIX 
(see Fig.~\ref{fig:comp_sol})
are fully consistent with those reported in 
\cite{BDD95}.
Moreover, since that paper provided also a seminalytical description 
of the solution of the 
Kompaneets equation, 
it is clear that a good agreement of the numerical results 
obtained with the original and update code represents
a further confirmation of the validity of the analytical 
description.\\

\subsubsection{Electron temperature behavior}
\label{sec:test_phi}

While the energy exchange between matter and radiation 
is leaded by Compton scattering
(in the absence of external energy dissipation processes), 
electrons reach the equilibrium Compton temperature 
\citep{peyraud, ZL70}, 
$T_{e,eq}$,
in a time shorter than the expansion time. For $y_h \ll 1$,
(late heating processes) Compton scattering is no longer able to 
significantly modify
the shape of the perturbed spectrum and the final electronic temperature $\phi_f$ (remember 
that $\phi = T_e/T_r$), immediately after the decoupling, is very close 
to $\phi_{eq} = T_{e,eq}/T_r$.
On the other hand, for energy injections at redshifts corresponding 
to $y_h \gtrsim 5$,
Compton scattering can establish kinetic equilibrium between matter and radiation. This
corresponds to a Bose-Einstein spectrum, 
with a final electron temperature given by
\citep{SZ70, DD77}:
\begin{equation}
\phi_f(y_h \gtrsim 5) = \phi_{BE} \simeq (1 - 1.11 \mu_0)^{-1/4}\,,
\label{eq_phi_test}
\end{equation}
where $\mu_0(\ll 1)$ is the usual dimensionless chemical 
potential. Moreover, in this case
the evolution of the chemical potential, $\mu(z)$, the relation between it and the
amount of fractional energy injected, $\Delta\epsilon/\epsilon_i$, depends on the energy 
injection epoch.

For the intermediate energy injection epochs, corresponding to $y_h \lesssim 5$, 
the final value of $\phi$ (a function depending on $y_h$) is between the values of 
$\phi_{BE}$ and $\phi_{eq}$, because the Compton
scattering works to produce a Bose-Einstein like spectrum 
anyway \citep{BDD91a}.
At these epochs the relation between the chemical potential and the amount of
fractional injected energy injected is simply given by 
\citep{SZ70, DD77}:
\begin{equation}
\mu_0 \simeq 1.4\,\frac{\Delta \epsilon}{\epsilon_i}\, .
\label{eq_mu0}
\end{equation}
By exploiting the numerical results,
a simple formula for $\phi$ can be found 
\citep{BDD95}:
\begin{equation}
\phi_f(y_h) = \frac{k}{5}\frac{5 - y_h}{k + y_h}(\phi_{eq} - \phi_{BE}) 
+ \phi_{BE}\, ;
\label{fiana}
\end{equation}
here $k=0.146$ is a constant derived from the fit. 
Moreover, this expression represents an 
accurate description of the
evolution of $\phi$ for any value of $y_h$. In facts, for a value of $y$ 
($y < y_h$) they found:
\begin{equation}
\phi(y,y_h) = \phi_f(y_h - y)\,.
\label{fievol}
\end{equation}
In the considered cases, as in many situations of interest, 
the perturbed spectrum of the radiation
(immediately after the heating process) is described by a superposition
 of blackbodies
and the equilibrium temperature is given by 
\citep{ZS69, ZEL72, BDD95}:
\begin{equation}
\phi_{eq} \simeq (1 + 5.4\,u) \phi_i\,,
\label{eq_phi_test2}
\end{equation}
where 
$\phi_i = T_i/T_r = (1 + \Delta\epsilon/\epsilon_i)^{-1/4} \simeq 1 - u$ 
and the
Comptonization parameter $u$ could be related to the amount of 
fractional energy exchanged by
\citep{ZS69, ZEL72, BDD95}:
\begin{equation}
u \simeq (1/4)\Delta\epsilon/\epsilon_i\,.
\label{eq_y_epsilon}
\end{equation}

\begin{figure}
 \hskip -0.5cm
 \includegraphics[width=1.1\hsize]{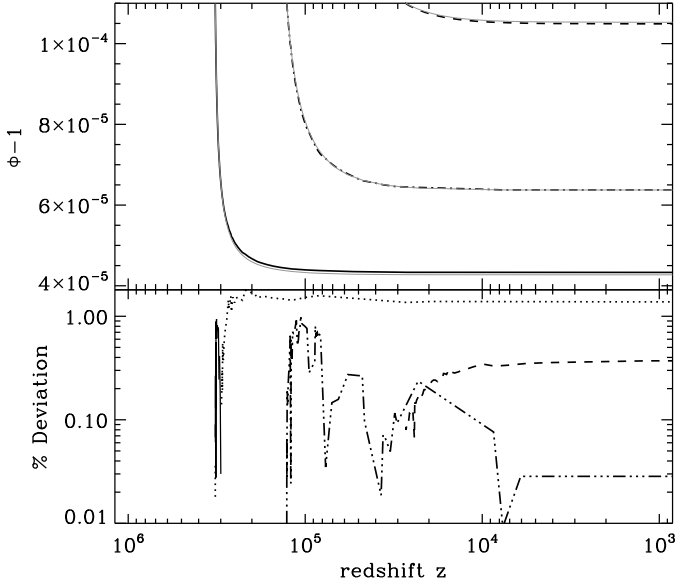}
 \bigskip
 \caption{Top panel: evolution of $\phi$ as derived from the numerical code.
The parameters adopted for the computation
are the same as in Fig.~\ref{fig:energy_cons} 
as well as the adopted kinds of lines. 
We report also for comparison the analytical results obtained from 
Eq.~(\ref{fiana}) (thin solid lines).
Bottom panel: relative differences between the
analytical results and the numerical results
expressed in terms of 
$(\phi_{analytical}-\phi_{numerical})/(\phi_{numerical}-1)$. 
Note that in absolute value they are $\lesssim 1\%$ at each time. 
Again, the adopted kinds of lines are as 
in Fig.~\ref{fig:energy_cons}, but solid line is replaced by dots
when the above relative difference is negative.
}
\label{fig:fievol}
\end{figure}

Eqs.~(\ref{fiana}) and (\ref{fievol}) can be used to test the 
behavior of the electron temperature during the
numerical integration of the Kompaneets equation carried 
out with the new code version. 
With the increasing of the time variable,
the values of $\phi$ are saved into the file DATIP, from the initial time step to the final one.
The upper panel of Fig.~\ref{fig:fievol} shows the two behaviors of 
$\phi$ (the numerical one and the
analytical expression given by Eqs.~(\ref{fiana}) and (\ref{fievol})) 
while the bottom panel displays 
their relative difference.

This test is of particular importance for the check of the validity 
of the results
because of the crucial role of $\phi$ in the Kompaneets equation.
As remembered in the previous section, 
in the new version of the code a different scheme is 
used with respect to that implemented in the original version.
The verification of the very good agreement of the above behaviors of $\phi$  
further supports the substantial equivalence of the two code versions 
and their reliability.
In particular, it confirms that 
the new adopted numerical scheme for the evolution of $\phi$, 
in principle less stable than the implicit scheme
implemented in the code original version, does not affect 
the validity of the solution. 

\subsection{Free-free distortion}
\label{sec:freefree}

As already discussed by \cite{SZ70},
accurate measures of the CMB spectrum
in the Rayleigh-Jeans region could provide quantitative informations about the 
thermal 
history of the Universe at primordial cosmic epochs. 
On the other hand, photon production processes (mainly radiative Compton at earlier 
epochs and 
bremsstrahlung at later epochs) work to reduce the CMB spectrum depression at long 
wavelengths (see \cite{DD80})
since they try to establish a true (Planckian) equilibrium. 
For $z < z_p$ \citep{DD80, BDD91a} with
\begin{equation}
z_p \simeq 2.14 \times 10^4 \bigg(\frac{T_0}{2.7\,K}\bigg)^{1/2}\bigg(\frac{k}
{1.68}\bigg)^{1/4}
\widehat{\Omega}_b^{-1/2}\,,
\label{eq_zp}
\end{equation}
low frequency photons are absorbed before Compton scattering 
moves them to higher frequencies.

In case of small and late distortions ($z_h \lesssim z_p$), a 
quite good approximation of the
whole spectrum is given by \citep{BDD95}
$$
\eta(x,\tau) = \eta_i e^{-(\tau-\tau_h)} + e^{-(\tau - \tau_h)} 
\int_{\tau_h}^{\tau}e^{(\tau' - \tau_h)}
\frac{1}{e^{x/\phi(\tau')} - 1}\textrm{d}\tau'
$$
\begin{equation}
\qquad + u \frac{x/\phi_i e^{x/\phi_i}}{(e^{x/\phi_i} - 1)^2}
\bigg(\frac{x/\phi_i}{\tanh (x/2\phi_i)} - 4\bigg)\,,
\label{eq_eta_approx}
\end{equation}
where the index $i$ denotes the initial value of the 
corresponding quantity and $u$ is the Comptonization parameter. This expression 
provides also an exhaustive description of continuum spectral distortion
generated in various scenarios of (standard or late) recombination
or associated to the cosmological reionization.
For an initial blackbody spectrum, at dimensionless frequencies
$x_B \ll x \ll 1$ the above equation can be simplified to
\citep{BDD95}
\begin{equation}
\eta \simeq \eta_{BB,i} + \frac{y_B}{x^3} - u\frac{2}{x/\phi_i}\,;
\end{equation}
here $y_B$, an optical depth of the Universe for
bremsstrahlung absorption (radiative Compton can be neglected at late epochs), 
is analogous to the Comptonization 
parameter and it is given by
\begin{equation}
y_B = \int_{t_h}^t (\phi - \phi_i)\phi^{-3/2}g_B(x,\phi)K_{0B}\textrm{d}t =
\end{equation}
$$
\int_{1+z}^{1+z_h}(\phi - \phi_i)\phi^{-3/2}
g_B(x,\phi)K_{0B}t_{exp}\frac{\textrm{d}(1+z)}{1+z}\,;
\label{eq_yb}
$$
$x_B$ is the frequency at which $y_{abs,B} = 1$
\citep{ZEL72, DEZ86}.
The dependence of the Gaunt factor 
\citep{KL61, RL79, BDD91a}
on $x$ and $\phi$ at very long wavelengths is weak: 
$g_B \propto \textrm{ln}(x/\phi)$.

In terms of brightness temperature, the distortions at low frequencies 
(at any redshift) can be written as
\begin{equation}
\frac{T_{br} - T_r \phi_i}{T_r} \simeq \frac{y_B}{x^2} - 2u\phi_i\, ,
\label{eq_tbr}
\end{equation}
where $T_r=T_0(1+z)$. This approximation holds
at low frequencies but not at too low frequencies, where
the brightness temperature obviously approaches the electron 
temperature because of the extremely high efficiency of bremsstrahlung,
still able to generate a Planckian spectrum at electron temperature.

\begin{figure}[h!]
 \hskip -0.4cm
 \includegraphics[scale=0.54]{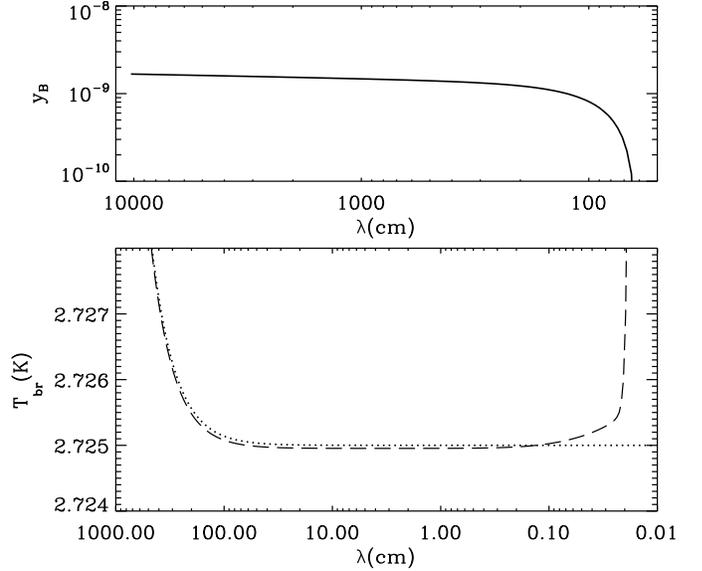}
 \bigskip
 \bigskip
 \caption{Top panel: $y_B$ as derived from Eq.~(\ref{eq_yb_num}). 
Bottom panel: comparison between the numerical (long dashes) solution 
and the analytical
approximation represented by Eq.~(\ref{eq:Tbry}).
See the text for further details.}
\label{fig:yb}
\end{figure}

In order to show that our numerical solution follows the behavior 
described by the above equation, we can compute
$y_B$ from the brightness temperature derived from the numerical solution:
\begin{equation}
y_B \simeq x^2\bigg(\frac{T_{br} -T_r \phi_i}{T_r} +2u\phi_i\bigg)\,.
\label{eq_yb_num}
\end{equation}
The reported numerical result (see Fig.~\ref{fig:yb}) refers to a 
heating process corresponding to
a full reionization starting at $z \simeq 20$ with 
$\phi = 10 \times {10^4}$ K, producing a final Comptonization 
parameter $u \simeq 4 \times 10^{-6}$ compatible with FIRAS upper 
limits. 
As shown in Fig.~\ref{fig:yb},
at low frequencies $y_B$ approaches an almost 
constant value.
This is correct only for $\lambda > 200 - 300$ cm
while at higher frequencies
$y_B$ is no longer almost constant (see again Fig.~\ref{fig:yb}) because 
of the
dependence of the Gaunt factor on $x$ and $\phi$ as 
expressed in the definition of $y_B$.
We can also write an expression describing the brightness temperature 
through a constant parameter $\bar{y}_B$, derived from
Eq.~(\ref{eq_yb_num}) (see Fig.~\ref{fig:yb})
averaged over the range at very long frequencies before the $y_B$
declining shown in Fig.~\ref{fig:yb}, 
to verify the accuracy of the below expression 
\begin{equation}
T_{br,y_B} = \bigg(\frac{\bar{y}_B}{x^2} -2u\phi_i + \phi_i\bigg) \cdot 
T_r 
\label{eq:Tbry}
\end{equation}
in comparison with the numerical results. 
Note that where the Gaunt factor dependence on $x$ and $\phi$ 
produces a significantly varying $y_B$ (see Fig.~\ref{fig:yb}), the 
Comptonization 
decrement is more relevant than the free-free excess in determining 
the brightness temperature, as evident from the good agreement of the two
curves in Fig.~\ref{fig:yb}. 
Clearly, the brightness temperature derived in this way works 
only up to frequencies ($\sim 100$~GHz)
approaching that at which the excess in the brightness 
temperature produced by the Comptonization begins (at $\sim 220$~GHz, see 
Fig.~\ref{fig:yb}). 

Note also that the use of $\bar{y}_B$ in 
Eq.~(\ref{eq:Tbry}) 
implies a slight excess 
($\sim ({y}_B-\bar{y}_B) T_r/x^2$)
with respect to the accurate numerical results 
since $y_B$ decreases with the frequency (see Fig.~\ref{fig:yb}). 
In this representative test the excess is $\approx 0.1$~mK, but 
obviously its value depends on the amplitude of free-free distortion
(other than on the frequency). 
This error is clearly negligible for the analysis of current data (see e.g.
\cite{SB02a,SB02b,TRIS,ARCADE}). 
It is likely not so relevant for the 
analysis of future measures at $\lambda \ga 1$~cm with 
accuracy comparable to that proposed for DIMES \citep{KOG96, BS03}.
On the contrary, it could be relevant 
for a very accurate analysis of
future measures at with precision comparable to that
proposed for FIRAS~II $\lambda \la 1$~cm \citep{FM02,BSZ,mather09}
or thought in ideas for possible future long wavelength experiments 
from the 
Moon$^{12,13}$\footnote{$^{12}$http://www.lnf.infn.it/conference/moon07/Program.html}
\footnote{$^{13}$http://sci.esa.int/science-e/www/object/index.cfm?fobjectid=40925} 
\citep{Moon}.
This calls for a complete frequency and thermal 
history dependent treatment of the free-free distortion 
in the accurate analysis of future data of extreme accuracy.

\section{Discussion and conclusion}
\label{sec:conc}

We have described the fundamental numerical approach 
and, in particular, the recent update to recent NAG versions
of a numerical code, KYPRIX, 
specifically written for the solution of the Kompaneets
equation in cosmological context, aimed to the very accurate computation
of the CMB spectral distortions
under quite general assumptions.
The recipes and tests 
described in this work can be useful to implement accurate 
numerical codes for other scientific purposes using 
the same or similar numerical libraries or to verify the validity 
of different codes aimed at the same or similar problems.

Specifically, we have discussed the main subdivisions of the code and
the most relevant aspects about technical specifications and code 
implementation. After a presentation of the 
equation formalism and of the boundary conditions added 
to the set of ordinary differential equations derived from
the original parabolic partial differential equation, 
we have given details on the adopted space (i.e. dimensionless frequency) 
grid, on the output results, on the accuracy parameters, and
on the used integration routines. The introduction
of the time dependence of the ratio between electron and photon 
temperatures 
and of the radiative Compton scattering term, both  
introducing integral terms in the Kompaneets equation, 
has been addressed in the specific context of the recent NAG versions
by discussing the solution adopted to solve the various related technical 
problems. 

Some of the tests we carried out to
verify the reliability, accuracy, and performance of the code
are presented.

We have compared the results of the update version of the code with
those obtained with the original one, reporting some representative 
cases, and we have found an excellent agreement.

Some specific quantitative tests are here reported. 
They indicate a very good accuracy in the energy conservation:
for appropriate choices of the code accuracy parameters, 
the fractional injected energy is conserved within an accuracy
better than 0.05\%, or, in other words, possible energy conservation
violations are negligible in practice for theoretical predictions
and for comparison with current and future data.
The time behavior of the electron temperature is found 
to be in excellent agreement with the results obtained 
with the original code version, in spite of the different
schemes adopted to update the evolving electron temperature.
These are important verifications
that probe that the current implementation of the code KYPRIX 
circumvents the problem represented by the lost possibility of
internally adapting the solver of partial differential equation 
to make it directly able to include integrals of the solution vector
exactly at the current time step.
Also, the setting of the accuracy parameters 
of the solver 
is now less flexible.
Of course, these features imply a certain increase of complexity in 
the code implementation, usage, and in the setting of code parameters. 
In spite of these difficulties, thanks to the better treatment of 
various computational steps involving integrals of functions over
finite intervals, 
the new version of KYPRIX achieves a good accuracy even 
in the treatment of very small distortions.

We have described also the introduction of the cosmological constant
in the terms controlling the general expansion of the Universe
in agreement with the current concordance model,
of the relevant chemical abundances, and on the ionization history,
from recombination to cosmological reionization.

 \begin{figure}[th!]
 \includegraphics[scale=0.52]{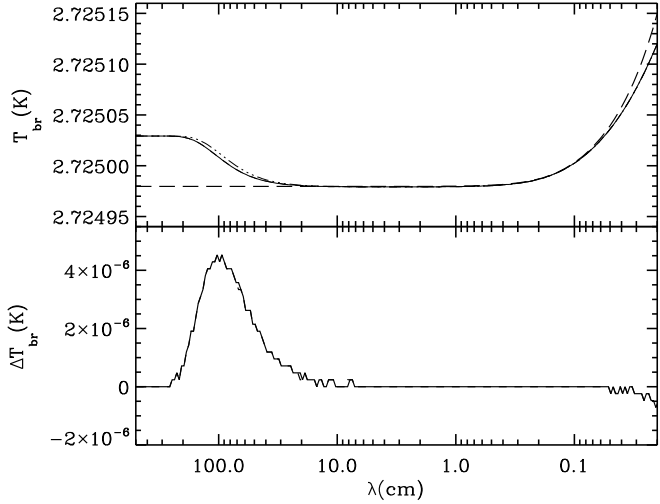}
 \bigskip
 \bigskip
 \caption{Final (i.e. at $z=1090$) distorted spectra for different 
treatments of the recombination process. 
We assume as initial condition a Comptonization 
spectrum at $z=10^4$ characterized by
$\Delta \epsilon_r/\epsilon_i = 10^{-5}$.
Top panel: initial condition (dashes), final spectrum 
in the case of instantaneous recombination at $z=1090$
following the fully ionized phase (three dots-dashes), final spectrum
in the case of gradual, more realistic modelizations of the 
recombination process (the two different treatments give 
results indistinguishable in these plots).
Bottom panel: difference between the final solution 
found in the case of instantaneous recombination and
of gradual, more modelizations.
See the text for further details.}
\label{fig:rec}
\end{figure}

 \begin{figure}[th!]
 \includegraphics[scale=0.52]{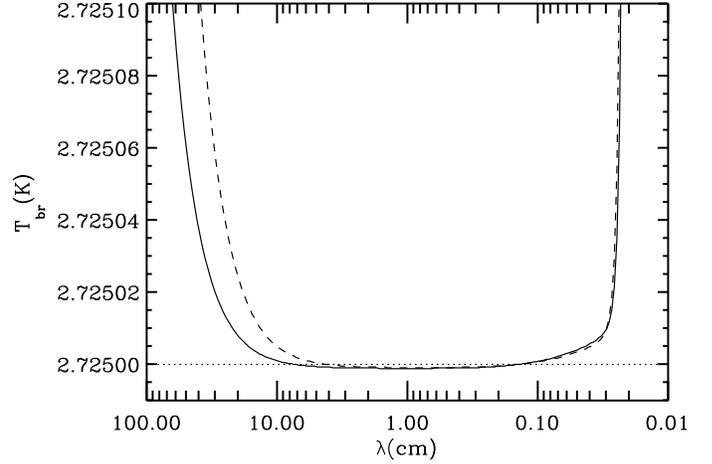}
 \smallskip
 \caption{Distorted spectra predicted for the considered reionization
models: initial blackbody spectrum (dots) at $z=20$, almost 
corresponding to the beginning of the heating phase 
in the considered model, 
final (at $z=0$) spectrum in the case of constant $\phi$~(=147) 
(dashes) and for the realistic implementation of the ionization 
and thermal history in the {\it suppression} model (solid line).
See the text for further details.}
\label{fig:reio}
\end{figure}

While an extensive discussion of various possible  
applications of these new features of the code, object of future work,
is not the scope of this paper,
we report here few representative examples.

In Fig.~\ref{fig:rec} we show the effect
of different treatments of the computation 
of the hydrogen and helium ionization fractions during the recombination 
epoch in the case of a relatively late spectral distortion.
Three different treatments of the recombination history 
are exploited: an instantaneous recombination following 
the fully ionized phase and two 
gradual, more realistic modelizations of the recombination process
characterized by different evolutions of ionization fractions.
In one of these two cases only the evolution of the electron ionization 
fraction, $\chi_e$, is taken from the code RECFAST
while the $H$ and $He$ ionization fractions 
are computed solving the system of Saha 
equations; in the other case we 
directly use the complete 
output of the code RECFAST
(see Sect.~\ref{sec:reion}
and the caption of Fig.~\ref{fig:rec} for more details).
The results obtained in these two cases are identical in practice,
while the assumption of full ionization up to 
$z=1090$ in the instantaneous recombination case
produces a certain overestimate ($\sim 10$\%) 
of the long wavelength excess due to (essentially free-free) 
photon production as well as a small amplification ($\sim 1$\%)
of the relaxation of the initial Comptonization spectrum
towards a Bose-Einstein like spectrum
(obviously ineffecient, at the considered redshifts).
With the adopted numerical accuracy  (ACC=$10^{-12}$) and grid points
($NPTS=30001$), the required CPU time on the used IBM platform is
of about 40 minutes, with less than 20\% differences between the 
three different cases.

Fig.~\ref{fig:reio} reports the results derived for a particular
astrophysical scenario of cosmological reionization. We adopt here the
evolution of the electron ionization fraction $\chi_e$
(the $H$ and $He$ ionization fractions
being then computed solving the system of Saha equations)
and of electron temperature predicted in the 
{\it suppression} model presented in \cite{schneider08}
and \cite{burigana08}
that implies a Thomson optical depth, $\tau \simeq 0.1017$,
almost in agreement with recent WMAP data (for consistency, 
we adopt in these computations
the same cosmological parameters used in the quoted works).
For comparison, we exploit a simple case 
with full ionization and a constant ratio, $\phi$,
between the electron temperature and the radiation temperature
which is chosen to have the same amount of 
energy injected in the plasma
($\Delta \epsilon_r/\epsilon_i \simeq 7.6 \times 10^{-7}$)
obtained in the above realistic model.
As expected by construction, 
the two very different histories produce
the same high frequency (Comptonization like) distortion
but the long wavelength region is very different 
because of the contribution of free-free 
emission$^{14}$\footnote{$^{14}$
In both cases, the density contrast in the 
intergalactic medium associated to the formation 
of cosmic structures has been not included.
The excess at low frequencies due to the free-free distortion
should be then considered as lower limit in both cases,
since the computation has been performed 
in the ``averaged density'' approximation.
Therefore, a correction factor 
$\approx$~$<$$n_e^2$$>/<$$n_e$$>^2$~$>1$,
coming from a proper inclusion of the treatment
of density contrast in the intergalactic medium,
should be applied to the free-free term.}.
The difference of this term in the two cases is 
mainly characterized by the evolution of the product
$\phi^{-1/2} \chi_e^2$ and by the fact that 
free-free is more efficient at higher redshifts 
in the simple case than in the realistic one
(see the caption of Fig.~\ref{fig:reio} for more details).
With the adopted numerical accuracy (ACC=$10^{-12}$) and grid points
($NPTS=30001$), the required CPU time on the used IBM platform is
of few hours for the realistic model and about one hour 
in the simple model with a constant $\phi$.

Finally, we have focussed on some properties of the free-free 
distortions, relevant for the long wavelength region of the CMB spectrum,
by checking that the new code version, as the original one, 
very accurately recovers the existing analytical approximations in their
limit of validity. We have also discussed the relevance of accurate
computations able to improve the simple treatment based on 
the approximation with a frequency independent free-free distortion 
parameter.

All the tests done demonstrates the reliability and 
versatility of the new code version and 
its very good accuracy and applicability to the scientific 
analysis of current CMB 
spectrum data and of those, much more precise, that will be 
available in the future. 

\section{Acknowledgments}

It is a pleasure to thank L. Danese, G. De Zotti, R. Salvaterra, and A. Zizzo 
for useful discussions and collaborations. 
Some of the calculations presented here have been
carried out on an alpha digital unix machine
at the IFP/CNR in Milano and at the IBM SP5/512 machine at CINECA-Bologna 
by using some NAG integration codes.
We warmly thank M. Genghini and C. Gheller for their kind assistance 
and technical support related to the machines used.
We acknowledge the use of the code RECFAST. 
We acknowledge the support by the ASI contract I/016/07/0 ``COFIS''.
We thank the anonymous Referee for constructive comments.

\end{document}